# Concerning the limit for neutrino mass

Akmarov A.A., Akmarov K.A.

Udmurt State University, the department of physics-energy, Izhevsk, Russia.

E-mail: akmarov@gmail.com

Abstract

The definition method of the limit for the neutrino mass of rest by using the analysis of the extreme values of the beta-particles distribution under beta-decay is given. The anomaly of distribution at the end of the curve was analyzed. The results of the experiments by spectrum analysis of electrons under tritium disintegration and the results recording of the stream of neutrinos from the supernova SN-1987A flash to estimate the limit for the neutrino mass were analyzed and compared. The problems of interpretation of the results obtained disappear in the case the mass less flavor of neutrino along with mass neutrino.

## Introduction

The question about neutrino mass – is one of the central questions in modern physics [1]. For the definition of neutrino properties, it is necessary to research the process in stars, including the Sun, to construct the evolution picture of Space. The definition of minimum mass of neutrino is a necessary step for understanding the beginning mass of elementary particles, the "dark matter" and "dark energy" Does a rest neutrino exist and what minimum mass has a massive neutrino? It took more than fifty years to attempt to make definitions of the neutrino mass of rest, but there is no convincing result.

## 1. The method for definition of the neutrino mass of rest using the curve of distribution the beta-particles by beta-decay

Fermi proposed experimental method for definition of neutrino mass of rest using maximum energy border of curve distribution of beta-particles by beta-decay. Nowadays, it is considered, that the best experimental result exceeded the use of this method was achieved by V.M.Lobashov's group in Troisk by means of research beta-decay of tritium [2]. They obtained sense of rest mass smaller then 2,2 eV. There is a large problem in doing this experiments, as it will be registered as maximum energy of electrons at beta-decay with precision as part of eV. The intensity of electrons stream is extremely small. Very much time is needed for experiment, more than one year of continuous work and a complicated instrumentation is needed for providing precision of order 1%. Some kinds of specified effects appear, that change the energy of target electrons. In particular, the appearance of "bamp" (it is a glitch of monochrome electrons at the end of spectral curve) forces the experimenters to invent artificial correction in order the value of the result was positive. The members of discussion [1] do not deny that for explanation of this effect a new physics may be developed.

Using extreme value, where the stream of electrons is a maximum, the authors [4] suggest a proposal variant of experiment for definition mass of rest of neutrino, on special plants, not on upper energy border of the curve distribution of electrons. Using the bond width of energy interval, limited by tangent to the curve in maximum point of calculated beta-specter of neutrino mass that equals zero and is bounded by points of intersection it with experimental curve for normalization specters and by antineutrino mass, which define the end point of specter (Fig.1). This method gives acceleration of statistic calculation.

The energy of electrons, in electron mass of rest is measured. The function, describing differential specter, must be normal in order to delete inessential factors. Then, the differential specter of beta-particles will be given by means of the equation:

$$N(\varepsilon, m_\nu) = (\varepsilon+1)\sqrt{\varepsilon(\varepsilon+2)}(\varepsilon_0-\varepsilon)\left[(\varepsilon_0-\varepsilon)^2 - m_\nu^2\right]^{1/2}$$

$$\varphi(m_\nu) = \int_0^{\varepsilon_0-m_\nu} (\varepsilon+1)\sqrt{\varepsilon(\varepsilon+2)}(\varepsilon_0-\varepsilon)\left[(\varepsilon_0-\varepsilon)^2 - m_\nu^2\right]^{1/2} d\varepsilon$$

where $\varepsilon = \dfrac{E}{m_e c^2}$, $\varepsilon_0 = \dfrac{E_0}{m_e c^2}$, $m_\nu = \dfrac{m_\nu}{m_e}$.

Altitude of maximum of curve, proper to $m_\nu = 0$, is determined from the equation:

$$\left.\frac{dN(\varepsilon,0)}{d\varepsilon}\right|_{\varepsilon=\varepsilon_m} = 0$$

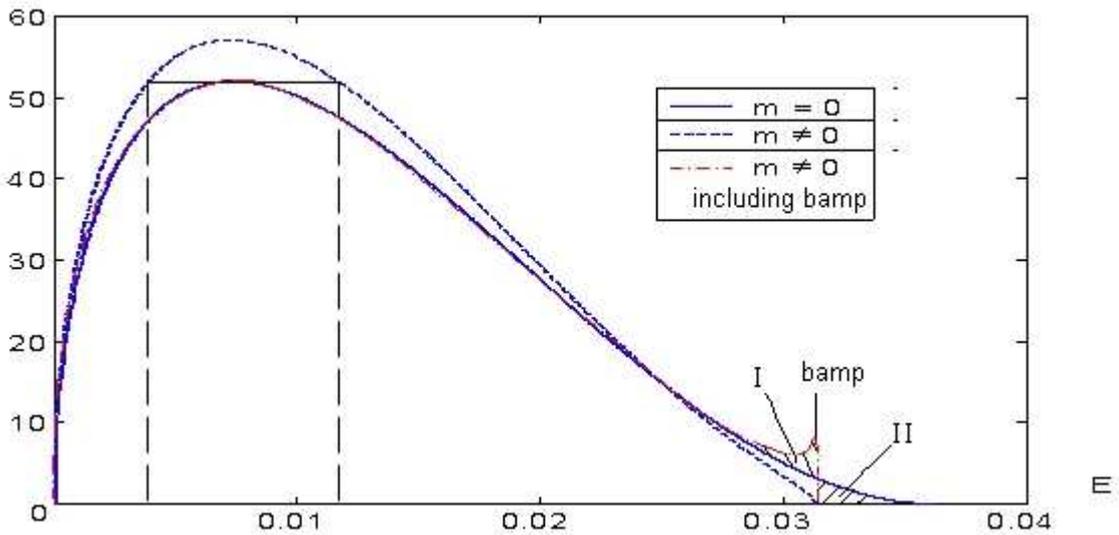

Fig.1. Curve of beta specter of tritium in case with the zero mass of antineutrino and non zero mass with "bamp" and without "bamp". The energy of the beta-particles denoted in electron mass of rest.

The width of energy interval equals the length of tangent $\Delta\varepsilon = \varepsilon_2 - \varepsilon_1$, where $\varepsilon_1$, $\varepsilon_2$ - are points of intersection of the tangent to curve, conform to $m_\nu = 0$ and curve of experimental specter. $\varepsilon_1$, $\varepsilon_2$ are determined from the equation:

$N(\varepsilon, m_\nu) = N(\varepsilon_m, 0)$,

having two roots $\varepsilon_1$, $\varepsilon_2$. So, the dependence of width on the energy interval on neutrino mass of rest, will be found. Under this condition the zero interval complies with zero mass.

If examining variant with "bamp", in maximum case, the curve of beta-specter will be almost congruent with curve by $m_\nu = 0$, except a little section near "bamp". After having normalized, the areas of the sections I and II (Fig.1) must be equal. This variant gives possibility to estimate minimum mass or energy parameters of antineutrino.

The "bamp" availability do not correspond modern conception of beta-decay and needs a new presentation. As "bamp" is located in a border of specter, where the electrons have maximum values of energy, and antineutrino minimum values of energy, both may be taken-off the atom of tritium. For it, extreme low energy antineutrino, but having magnetic moment, must overcome the energy field of the frame of electron, the potential ionization of tritium, equal – near 13,6 eV. As far as antineutrino cannot stay with in atom, it uses deficient part of the energy from electron, coming out with it. In this case the energy of the bordering electrons must be scale down to this value – 13,6 eV. Then the number of electrons with energy decreased to this 13,6 eV. will be increase. Experimentally, at that time "…any structure supervised in a kind of glut of intensity offset to 5 – 15 eV, from border point, in side low energy" [3]. It confirm adopted assumption. It follows, that the extreme low energy antineutrinos will not be able to overcome the frame of electron using tunnel mechanism. Or it is used only by the top of frame. The next fortune of its antineutrinos is unknown. May be it's added to the "dark matter". Then the number of "dark mater" shall be increase in the Space. In the second variant, without energy this particle can't exist having mass and spin at one time. We suppose that the mass and spin separated as a spin-charge separation at low energy of electrons [5]. The entropy of system will be increased more quickly if it goes out of here with light velocity mass less as a photon.

It is impossible to understand, from where the electrons by beta-decay with energy more, than energy of reaction, appear. It is interesting, that surplus energy of electrons equals about half of energy ionization of tritium.

## 2. Discuss about the mass rest of neutrino, using the results recording of the stream of neutrinos from the supernova SN-1987

In principle another method for definition of the limit of neutrino mass present the registration of neutrino stream from collapse of stars. It may be used for estimation of the limit of the neutrino mass, showed in [6].This possibility appeared in 1987, February, the flash was registered by supernova SN-1987A. It appears probably – once in about 30 -100 years. Two packages of neutrino occurrences were registered. The impulses were registered in different laboratories of the Earth. The calculation [7] of restriction to neutrino mass of rest was performed from suppositions, that if two neutrinos with energies $E_1$ and $E_2$ run out at the same time, at the and of the distance d one appears behind from another on the interval of time $\Delta t$. So, the neutrino mass of rest $m_\nu = E_1 E_2 (2c\Delta t)^{0,5} \{ d ( E_2^2 – E_1^2 ) \}^{-0,5}$, where c-is a velocity of light, $E_2 > E_1$. According to the results obtained by other authors lay in diapason from 4eV up to 24eV. It is supposed, that in this formula the impulses correspond neutrinos of the same sort. Is that so? Working at this moment the neutrino detectors may register the electrons neutrino mainly. The heavier neutrinos could oscillate according to effect of Mickheev-Smirnov-Wolfenshtein (MSW) in different sorts of neutrinos. The fraction of electrons neutrino may be registered in a common packet.

It is difficult for many others to describe these results. It happened because the experimental material is very small, the processes, in the collapse of stars, are not clear enough and the properties of neutrino has not been studied yet. It is surprising that some impulses from the first packet has been registered before the signal from the gravitational antenna. Another problem is the lagging of the second packet from the first packet on about 17000 seconds. This problem is not the solution using variants with μ-neutrinos or τ-neutrinos, because the detectors, working on the Earth at that moment, can register only the electron-neutrinos. The process of collapse of Stars is examined as two stages, separated by a large period of time, today there is no basis enough to do it.

In this work it is suggested, to perform the calculation according the impulse, received by the gravitational antenna in Rome. All signals were separated into two packets. The first packet will be separated into three parts. One of the parts passed ahead of the impulse a little from the gravitational antenna. The second part lagged to $\Delta t$ = 0-2 sec. The third part lagged from the gravitation signal to $\Delta t$ = 7-10 sec. The second packet of impulses lagged near to $\Delta t$ = 17000sec. The final optical registration was performed only several hours later. This registration may be interpreted as the registration of neutrinos of different flavors. A simple calculation is done, if the velocity of the gravitational signal equals the velocity of light, will be written down as: $\Delta t = d/V - d/c$, then $V = cd/(c\Delta t + d)$ and $(1 - V^2/c^2)^{0.5} \approx (2c d\Delta t)^{0.5}/d$; where $\Delta t$ – is a difference from the time of the gravitational impulse and the time from the neutrino impulse; d – 52kPs – is the distance from supernova to the Earth; V – velocity of neutrino. The neutrino energy is measured as 20 – 40 MeV. There are no satisfactory results, if these simple calculations, are used. Either energy is not enough or the neutrinos will fly to the Earth much more later.

It is astonishing, that the light signal from flash does not run down the neutrinos, over so much time. In that case, the velocity of neutrinos equals the velocity of light travelling through the Space, to consider to necessary.

Let's examine the beginning of collapse of Stars with possible mass-less neutrino. The blanket from Star contains electron neutrinos. Lower contains layer µ-neutrinos. Some more lower contains layer τ-neutrinos [9]. The density of them is large, because the Stars mass is very large. Let neutrino have a visible magnetic moment. It confirmed indirectly by the picture of the Sun from neutrinos [10]. The picture is more, than on optical picture. Therefore, the trajectory of neutrinos bends from magnetic fields toward the Sun. The calculated magnetic parameters, seem to be very large, but we do not know about this enough. The temperature on the blanket is very large, so the magnetic moment of neutrino makes electro-dynamic field. In that case the reaction $\gamma + \nu_e \to e^+ + e^- + \nu_{ef}$ is possible. Mass-less neutrinos from photon $\nu_{ef}$, leave the Star with light velocity. Electrons and positrons go away from the surface into the Star. Gradient of gravitational field increases. This process may lead to coherent evolution of neutrinos flavors [11]. And collapse will begin. Any impulses lagged from the gravitational impulse, but the optical signal does not catch up it. It is possible, the impulses belong to heavy neutrinos, which transform into mass-less flavor and fly across the Space with light velocity. This transformation may use the process of oscillation into electron neutrinos then into photon neutrinos. It is possible, probably to have reactions: $\gamma + \nu_\mu \to \mu^+ + \mu^- + \nu_{\mu f}$, and $\gamma + \nu_\tau \to \tau^+ + \tau^- + \nu_{\tau f}$ Lagging from the gravitational impulse may be explained as evaporation from the surface of the star. As a result, at first the electron neutrinos as photon neutrinos fly from the surface, and then the collapse begins with formation of gravitational impulse. Then, the evaporation with transformation into photon neutrinos goes for heavy neutrinos. The various photon neutrinos fly into Space with velocity of light. Inside the Earth they transform into massive neutrinos and oscillate into different flavors. The electron neutrinos fraction is registered by detectors. The registration on the Earth is done as follows: at first the registration of neutrino, then the gravitational signal, then again registration of neutrino and then many hours later the optic flash registration. Just like this, the order of registration is observed.

The results of tritium experiment and flash SN-1987A were compared with calculation; there is supposition of existing the neutrino with zero mass, at the same time with massive neutrinos.

# Conclusion

One can calculate experimental results of beta-decay using extreme values on the curve, where the stream of electrons is maximum.

The conduct anomalous of electrons with maximum energy in tritium experiment, gives arguments for supposition, which the neutrino with extremely low energy move to neutrinos with zero mass of rest.

The analysis of registration of the flash supernova SN-1987A, gives possibility to suppose that the collapse of the Star begins with transformation of the massive electron neutrinos into the mass of less modification. The balance upset between different sorts of neutrinos. The massive heavy neutrinos are transformed step by step, to more light sorts of neutrinos by MSW effect in the matter of star. The electrons and positrons fall inward the Star, increasing the gradient of gravitational field. The mass less neutrinos, such as photons, leaves the Star with a light velocity. The same process may be happened with heavy sorts of neutrinos, it may be difficult and compound.

All sorts of neutrinos go from the Space with light velocity as a photon without mass of rest. The neutrinos make oscillation of different sorts in the matter of the Earth by MSW-effect. The fractions of electron neutrinos will be registered by working detectors.